\documentclass[preprint,aps,showpacs]{revtex4}
 \usepackage{graphicx,amsfonts}
\usepackage{epsfig}
\usepackage{dcolumn}
\usepackage{bm}
\begin{document}

\preprint{IFT-P.020/2005}

\title{The Invisible Axion and Neutrino Masses}
\author{
Alex G. Dias\footnote{e-mail: alexdias@fma.if.usp.br}}
\affiliation{Instituto de F\'\i sica, Universidade de S\~ao Paulo, \\
C. P. 66.318, 05315-970 S\~ao Paulo, SP, Brazil. }

\author{V. Pleitez\footnote{e-mail: vicente@ift.unesp.br}}
\affiliation{Instituto de F\'\i sica, Te\'orica, Universidade Estadual
Paulista \\
Rua Pamplona 145, S\~ao Paulo, SP, Brazil. }
\date{\today}
\begin{abstract}
We show that in any invisible axion model due to the effects of effective 
non-renormalizable interactions related to an energy scale near the 
Peccei-Quinn, grand unification or even the Planck scale, active neutrinos 
necessarily acquire masses in the sub-eV range. Moreover, if  
sterile neutrinos are also included and if appropriate cyclic 
$Z_N$ symmetries are imposed, it is possible that some of these 
neutrinos are heavy while others are light. 
\end{abstract}
\pacs{14.80.Mz, 14.60.St, 14.60.Pq}
\maketitle

A natural and elegant way to explain the small value of the active
neutrino's masses is the so-called seesaw mechanism which is implemented
when \textit{heavy} right-handed sterile neutrinos, i. e., transforming as 
singlet under the standard model $SU(2)_L\otimes U(1)_Y$ gauge symmetry, are 
added~\cite{seesaw}. Moreover, depending on future neutrino oscillation
data, \textit{light} sterile neutrinos~\cite{lsnd} may be a necessary
ingredient of the physics beyond the standard model. 
From neutrino oscillation experiments we already know
that active neutrinos have a non-vanishing mass in the sub-eV 
region~\cite{pdg,nusosc} but, since the $Z^0$ invisible width implies 
the existence of  only three light active 
neutrinos, any additional light neutrino has to be sterile and there are 
several possibilities that keep consistency with LEP data~\cite{cj,tau2}. 
The problem is how to implement, in a natural way, light sterile neutrinos. 
Since they are not protected by the standard model symmetries, they may acquire
Majorana masses of the order of the next (if any) energy scale. A solution 
is the addition of sterile Higgs scalars and a new exact global 
symmetry~\cite{late,will}. The important point is that such a symmetry 
forbids the Dirac mass term avoiding in this way the seesaw mechanism and the 
sterile scalar singlet having a vacuum expectation value chosen just to 
generate light sterile neutrinos. Light sterile neutrino may also appear in
supersymmetric models~\cite{kang}. 

On the other hand, the introduction of a global chiral
(Peccei-Quinn) symmetry is an elegant solution to the strong CP
problem~\cite{pq} implying in the existence of a pseudo Goldstone boson, the
axion~\cite{axion}, which besides solving the strong CP problem it is certainly
a leading candidate for dark matter. Searches for the axion has been
done over the years. Recently, it was obtained an upper limit for the axion-photon
coupling $g_{a\gamma\gamma}<1.16\times10^{-10}\,\textrm{GeV}^{-1}$ if 
$m_a\lesssim0.02$ eV~\cite{cast,raffelt}. In fact, the expected mass for the axion,
coming from several experimental or observational constraints, is in the interval 
$10^{-6}<m_a<10^{-2}$ eV and we see that this interval is near to 
the neutrino masses required to explain solar and atmospheric neutrino 
data~\cite{pdg}. This fact suggests the existence of a 
common new energy scale as the responsible for such small masses, which in 
this case would be that one related to the invisible axion. 

Although the existence of a relation between the axion and the 
seesaw mechanism for generating neutrino masses has already been considered in
particular models~\cite{ms}, here we will put forward that it is inevitable that
neutrino get masses in any invisible axion model. Moreover, if the model have
also right-handed neutrinos, some of them may be light but the other ones may be 
heavy. In fact, we expect that if all sterile neutrinos have the same physical 
origin i. e., they are related to the same energy scale, it would be
natural that all of them are either light or heavy. However, if some of them are
light and the others heavy this may be seen as an evidence of 
energy scales different from the electroweak and Planck scales. 
The existence of these energy scales can be masqueraded by imposing discrete 
symmetries ($Z_N$'s) under which each neutrino can transform in a different way 
from each other.

The standard model (SM) an many of its extensions can be seen as effective 
theories below an energy scale $\Lambda$. The minimal model has doublets 
of left-handed quarks and leptons, the respective right-handed singlets for the
charged fermions,  and one doublet of Higgs scalars, denoted by $Q_L$, $\psi_L$, $u_R$, 
$d_R$, $l_R$ and $H$, respectively. Since axion models need more 
than one Higgs doublets, we will consider a model that
has at least two Higgs doublets. Next, we introduce $m\geq1$ 
right-handed neutrinos, $n_{sR}$, transforming as singlet 
under the SM gauge symmetries. An invisible axion model is one in which there 
is an approximate global symmetry protecting the $CP$ invariance of the QCD and 
which is break, beside by non-perturbative effects, by the VEV of an scalar singlet
$\phi$ added to the particle content with its imaginary part being almost the axion 
field~\cite{singleto}. Let us suppose also that the invisible axion is protected 
against gravitational effects by a local, in the sense of Ref.~\cite{kw}, 
$Z_N$'s cyclic symmetries~\cite{axion331,babu}. In general,
these symmetries can be anomalous but we will not address this issue here.
On the other hand, since the total lepton number is an accidental symmetry of 
the minimal SM without right-handed neutrinos, there are no reasons for this 
number to be conserved in extensions of the latter model, so we will also assume 
that the total lepton number can be violated explicitly in some non-renormalizable
interactions.

In the situation described above, if the Dirac mass term $\bar{\psi}_LHn_R$
is forbidden by the a $Z_N$ symmetry, there is always an effective operator,
\begin{equation}
-{\cal L}_D= \frac{f_{as}}{\Lambda^{N_\phi}}\,
\overline{\psi_{iaL}}\epsilon_{ik}H_kn_{sR}\phi^{N_\phi}+H.c.,
\label{dirac}
\end{equation}
with $N_\phi\geq1$ and $a$ is generation index, while $i,k$ refer to $SU(2)$ 
indices, the number of right-handed neutrinos is $s=1,\cdots,m$ and
we have omitted summation symbols. 
This term generates a Dirac mass when both $H_k$ and $\phi$ get 
a non-zero vacuum expectation value, 
$\langle \phi\rangle\approx\Lambda_{PQ}$, $\langle H_k\rangle$ 
is of the order of the electroweak scale ($\sqrt{\small{\Sigma}_k\langle H_k\rangle^2}\sim246$ 
GeV). Notice that this Dirac mass from Eq.~(\ref{dirac}) is proportional to 
\begin{equation}
{\cal M}_D= f \left(\frac{\Lambda_{PQ}}{\Lambda}\right)^{N_\phi}\,\langle
H\rangle,
\label{diracmass}
\end{equation}
so it may be small, in the sub-eV region even if $f\sim O(1)$, if 
$\Lambda\gg\Lambda_{PQ}$ and depending also on the $Z_N$ symmetry. However, the 
dimensionless matrix elements denoted by $f,f^\prime$ etc, may include loop
suppression factors.
 
In general, we may have also the $d=5$ interaction~\cite{sw79}, proportional to
$(1/\Lambda)\,\overline{(\psi_{aL})^c}\psi_{bL}HH$, that induces a Majorana mass
for the left-handed neutrino. However, if this interaction is forbidden by
$Z_N$'s there are effective interactions like, 
\begin{equation}
-{\cal L}_L=\frac{f'_{ab}}{\Lambda^{N^\prime_\phi+1}}\,\overline{(\psi_{ai})^c_L}
\epsilon_{ik} 
\psi_{bpL}\epsilon_{pl}H_kH_l \,\phi^{N^\prime_\phi}
+H.c.,
\label{teoa}
\end{equation}
with $i,k,p,l$ are $SU(2)$ indices, that are allowed with an appropriate
$N^\prime_\phi\geq1$. When $\phi$ (may be $\phi^*$) and $H$ gain a non-zero
vacuum expectation value they induces a Majorana mass to the left-handed
neutrinos  
\begin{equation}
{\cal M}_L=f^\prime\,\left(\frac{\langle
H\rangle^2}{\Lambda}\right)\left(\frac{\Lambda_{PQ}}{\Lambda}
\right)^{N^\prime_\phi},
\label{majorana1} 
\end{equation}
which is in the sub-eV range, for a given value of $\Lambda_{PQ}$, 
depending on the value of $\Lambda$ and $N^\prime_\phi$ and without 
any fine tuning in the dimensionless $f^\prime$ parameters.

Next, we note that the tree level mass term proportional to
$\overline{(n_{sR})^c}n_{tR}\phi$ with $s,t~=~1,\cdots, m$, 
induces a large Majorana mass 
for the right-handed neutrinos, if it is allowed at the tree 
level~\cite{ms}. However, if this mass term is also forbidden by 
the $Z_N$ symmetries, the effective interactions with lower dimension 
are
\begin{equation}
-{\cal L}_R=
\frac{f^{''}_{st}}
{\Lambda^{N^{\prime\prime}_\phi}}\,\overline{(n_{sR})^c}
n_{tR}\,
\phi^{N^{\prime\prime}_\phi+1}+ \frac{f^H_{st}}
{\Lambda^{2N_H-1}}
\,\overline{(n_{sR})^c}n_{tR}\,(H^\dagger_1 H_2)^{N_H} +H.c., 
\label{teob}
\end{equation}
will always be possible for $N^{\prime\prime}_\phi,N_H\geq1$ (here 
also $\phi$ may be $\phi^*$). The first term in Eq.~(\ref{teob}) 
induce a large Majorana mass if $\Lambda=\Lambda_{PQ}$, however 
this is not necessarily the case if $\Lambda>\Lambda_{PQ}$ and 
an appropriate $Z_N$ is introduced (see below).
The second term in Eq.~(\ref{teob}) arises because axion models 
need at least two electroweak doublets, say $H_1$ and $H_2$, and 
it also implies small Majorana masses for the sterile neutrinos, 
for instance when $N_H=1$, and $\Lambda\geq\Lambda_{PQ}$. 

As an illustration, we consider the first term in Eq.~(\ref{teob}) 
which generates the mass term 
\begin{equation}
{\cal M}_R=f^{\prime\prime}\,\left[
\frac{\Lambda_{PQ}}{10^{12}\,\textrm{GeV}}\,
\left(
\frac{\Lambda_{PQ}}{\Lambda}\right)^{N^{\prime\prime}_\phi}\right]10^{21}\;
\textrm{eV}.
\label{ef} 
\label{gutgra}
\end{equation}
As we said before, in Eqs.~(\ref{dirac}), (\ref{teoa}) and (\ref{teob}), 
$\Lambda$ is related to the new physics implying the non-renormalizable
interactions. It may be related to the PQ, grand unification (GUT) or even the
Planck scale~\cite{rb,kang}. 
We see from Eq.~(\ref{ef}) that, when $\Lambda=\Lambda_{PQ}$ (or
$N^{\prime\prime}_\phi=0$), the right-handed
neutrinos are necessarily heavy. We obtain 
${\cal M}_R=f^{\prime\prime}(10^{-4N^{\prime\prime}_\phi})10^{21}$ eV
if the $\Lambda$ is the GUT scale, i. e., $\Lambda=10^{16}$ GeV; or ${\cal
M}_R=f^{\prime\prime}(10^{-7N^{\prime\prime}_\phi})10^{21}$ eV if 
$\Lambda$ is the Planck scale, with $\Lambda_{PQ}=10^{12}$ GeV in both cases. 
We see that there exist a $N^{\prime\prime}_\phi$
which always produce neutrino masses of the order of eV. For instance, if we
assume that $f^{\prime\prime}\sim O(1)$, we have that $N^{\prime\prime}_\phi=5$
(GUT scale) or $N^{\prime\prime}_\phi=3$ (Planck scale) given in fact ${\cal
M}_R$ in the eV range. These value for $N^{\prime\prime}_\phi$ implies an
appropriate $Z_N$ symmetry. The important point is that
because of this symmetry some sterile neutrinos are
light but other are heavy, depending how they transform under
$Z_N$.

The general mass matrix for the neutrinos is
\begin{equation}
M^\nu=\left(
\begin{array}{ll}
{\cal M}_L & {\cal M}_D \\
{\cal M}^T_D & {\cal M}_R
\end{array}
\right).
\label{mg}
\end{equation}
If there is a hierarchy like $M_L\ll {\cal M}_D\ll M_R$ the eigenvalues of
such a matrix are of the form ${\cal M}_R$ and $-({\cal M}_L+{\cal M}^2_D/{\cal
M}_R)$. If the Dirac mass is not suppressed with respect to the
left- and right-handed Majorana mass terms, i. e., if
$N_\phi\ll N^\prime_\phi, N^{\prime\prime}_\phi$, we have that
neutrinos are pseudo-Dirac particles ${\cal M}_L,{\cal M}_R\ll {\cal
M}_D$~\cite{lw}. Notice that, as we said before, since there are 
several sterile neutrinos, they may not have all the same 
$Z_N$ charge, thus some may have the interaction in Eq.~(\ref{teob}) with
$N^{\prime\prime}_\phi=1$ and are heavy, but other get the same interactions
with $N^{\prime\prime}_\phi>1$ and are light. The exact value depend on the
value of the scale $\Lambda$. 
Hence, some entries of the matrix ${\cal M}_R$ may
be large while other may be small, implementing in this way the seesaw mechanism
for the active neutrinos and at the same time allowing light sterile neutrinos.
 
Let us considered an example of the present 
mechanism for generating both light and heavy sterile neutrinos in the 
context of a invisible axion model which is a version of the model of 
Ref.~\cite{axionsm}. The representation content is, in the fermion sector,
with three generations, lepton doublets $\psi$, quark doublets $Q$, the 
respective right-handed singlets $u_R,d_R$, $l_R$ and $\nu_R$. In the scalar sector,
there are four scalar doublets $H_u,H_d,H_l,H_\nu$, a scalar non-hermitian triplet
$T$. We avoid the scalar singlet $h^+$ and introduce a fourth right-handed neutrino,
$n_{4R}$. The cyclic symmetry is $Z_{13}$ ($\omega_k=e^{2\pi ik/13};\,k=0,\cdots,6$), 
but now fields transform as: $\psi\to\omega_6\psi$, $l_R\to\omega_4l_R$, 
$Q\to \omega_5Q$, $u_R\to\omega_3u_R$, $d_R\to\omega^{-1}_5d_R$ and 
$\nu_{aR}\to\omega_1\nu_{aR}$ 
$(a=e,\mu,\tau)$, $H_u\to\omega^{-1}_2H_u$, $H_d\to\omega^{-1}_3H_d$, 
$H_l\to\omega_2H_l$, and $H_\nu\to\omega^{-1}_6H_\nu$, $T\to\omega^{-1}_4T$, 
$\phi\to\omega^{-1}_1\phi$, $n_{4R}\to\omega^{-1}_4n_{4R}$, while the gauge 
fields transforming trivially. Notice that in this case it is not necessary to
introduce a $Z_3$ cyclic symmetry as it was done in Ref.~\cite{axionsm}. 
With these fields we have several effective interactions, the dominant are the 
following:
\begin{eqnarray}
-{\cal L}_Y&=&\frac{f_{ab}}{\Lambda}\,\overline{\psi_{aL}}\nu_{bR}\tilde{H}_\nu
\phi+\frac{f'_{ab}}{\Lambda}\,\overline{(\psi_{ai})^c_L}
\epsilon_{ik} (\psi_{bp})_L\epsilon_{pl}(H_\nu)_k(H_\nu)_l +
\frac{f^{\prime\prime}_{ab}}{\Lambda} \overline{(\nu_{aR})^c}\,\nu_{bR}\phi^2
\nonumber \\ &+& 
\frac{f^{\prime\prime}_{44}}{\Lambda^4}\overline{(n_{4R})^c}\,n_{4R}\phi^{*5}
+
\frac{h_{a4}}{\Lambda^4}\overline{n_{4R}}\,\nu_{aR}\phi^5
+H.c.,
\label{yukawa}
\end{eqnarray}
where where $\tilde{H}_\nu=\epsilon H^*_\nu$. These interactions
imply 
\begin{equation}
m_D\sim f_{ab}\langle H_\nu\rangle \frac{\Lambda_{PQ}}{\Lambda},\, m_L
\sim f'_{ab}\,\frac{\langle H_\nu\rangle^2}{\Lambda},\,
m_R\sim f^{\prime\prime}_{ab}\frac{\Lambda^2_{PQ}}{\Lambda},\, 
m_{44}\sim f^{\prime\prime}_{44}\frac{\Lambda^5_{PQ}}{\Lambda^4},\,  
m_{4R}\sim h_{a4}\frac{\Lambda^5_{PQ}}{\Lambda^4}.
\label{ora}
\end{equation}
The values for these entries of the neutrino mass matrix depend on the actual 
value of $\langle H_\nu\rangle$, $\Lambda_{PQ}$ and $\Lambda$. Let us suppose, 
just for illustration that $\langle H_\nu\rangle=100$ GeV, $\Lambda_{PQ}=10^{12}$ 
GeV and $\Lambda=10^{19}$ GeV. In this case we have 
$m_D\sim 10^4\,f_{ab}$ eV, $m_L\sim 10^{-6}\, f^\prime_{ab}$~eV, 
$m_R\sim f^{\prime\prime}_{ab}10^5$~GeV,
$m_{44}\sim 10^{-7}\,f^{\prime\prime}_{44}$ eV, and 
$m_{4a}\sim 10^{-7} h_{4a}$ eV. We see that the general matrix as in 
Eq.~(\ref{mg}), in this particular case, 
allows a light sterile neutrino which is most $n_{4R}$ while $\nu_{aR}$ are 
heavy ones. If $\Lambda_{PQ}=10^9$ GeV and $\Lambda=
\Lambda_{GUT}=10^{16}$ GeV we have: $m_D\sim 10^4\,f_{ab}$ eV, $m_L\sim 
10^{-5}\,f^\prime_{ab}$~eV, $m_R\sim 100\, f^{\prime\prime}_{ab}$ GeV, 
$m_{44}\sim 10^{-10}\,f_{44}$ eV and $m_{4R}\sim 10^{-10}\, h_{4a}$ eV. 
The scalar Higgs sector in this model is the same (up to the singlet $h^+$) 
with that of Ref.~\cite{axionsm}, and there it was shown that the scalar potential
with this particle content is consistent with the discrete $Z_{13}$ symmetry. 

The above mechanism is also implemented in models in which the main part of the
axion field is in a non-trivial representation. 
For instance the $SU(5)$ invisible axion model~\cite{su5} in which the 
axion is primarily the antisymmetric part of the singlet component of 
a complex $\textbf{24}$, $\Sigma$. 
If right-handed neutrino singlets are introduced the 
effective interaction  
\begin{equation}
{\cal O}^{\,\prime R}_{st}\propto\frac{1}{\Lambda}\;\overline{(n_{sR})^c}
n_{tR}\;\Sigma^*\Sigma,
\label{su5}
\end{equation}
implies heavy right-handed sterile neutrinos. Since $\langle\Sigma\rangle$ 
breaks down also $SU(5)\to SU(3)\otimes SU(2)\otimes U(1)$, the
unification scale is of the order of the PQ scale. It is interesting that in the
axion model considered above~\cite{axionsm}, the PQ symmetry is automatic, the
axion is protected from gravitational effects and unification occurs near the 
PQ scale having still an stable proton~\cite{321run}. This is just an illustration 
of how the issues of grand unification, the axion and the generation of neutrino 
masses can be related to each other. It is interesting to note that if it
does not exist any energy scale between the electroweak and the Planck scale,
with only active neutrinos, it would be difficult to generate neutrino masses
of the order of eV since in this case the only possible effective 
operator is $(1/\Lambda_{Planck})\,\overline{(\psi_{aL})^c}\psi_{bL}HH$ and 
the suppression factor is too large.  

Finally, a remark is in order. Since we are considering generic invisible 
axion models the couplings of the axion with all fermions, but the heavy 
sterile neutrinos, are strongly suppressed. It means that there is no 
potential conflict of these mechanisms for ge\-ne\-ra\-ting neutrino masses 
neither with big-bang nucleosynthesis nor with the cosmic microwave 
background~\cite{late,cline}. However these issues deserve a more carefully 
analysis because such constraints are highly model dependent.   

Summarizing, we have shown that an invisible axion implies light massive
active neutrinos. Furthermore, if the model has also several right-handed 
sterile neutrinos, some of them get large and others small Majorana masses, 
depending on how they transform under discrete $Z_N$ symmetries. 
The seesaw mechanism is implemented by the heavy neutrinos and, since 
there may be light sterile neutrinos, some neutrinos
may be pseudo-Dirac particles.

\end{document}